\documentclass[12pt,preprint]{aastex}

\usepackage{amssymb}
\usepackage{amsmath}
\usepackage{graphicx}
\begin{document}

\title{Polarization of dust emission in clumpy molecular clouds and cores}

\author{T.J. Bethell, A. Chepurnov, A. Lazarian \& J. Kim}

\newcommand{\mj}{J_{V}/I_{V}^{0}}

\newcommand{\av}{A_{V}}

\newcommand{\tg}{T_{g}}

\newcommand{\jlamat}{j_{\lambda}(a,T_{g})}

\newcommand{\omrad}{\Omega_{rad}}

\newcommand{\omt}{\Omega_{th}}

\newcommand{\jlamx}{J_{\lambda}(\mathbf{x})}

\newcommand{\glamx}{\gamma_{\lambda}(\mathbf{x})}

\newcommand{\gx}{\gamma(\mathbf{x})}

\newcommand{\ilamkx}{I_{\lambda}(\mathbf{k},\mathbf{x})}

\newcommand{\kb}{\mathbf{k}}

\newcommand{\xb}{\mathbf{x}}

\newcommand{\nh}{n_{H}}

\newcommand{\pvec}{\mathbf{P}}

\newcommand{\bvec}{\mathbf{B_{proj}}}

\begin{abstract}
Grain alignment theory has reached the stage where quantitative predictions
of the degree of alignment and its variations with optical depth
are possible. With the goal of studying the
effect of clumpiness on the sub-millimeter and far infrared
polarization we computed the polarization due to alignment via radiative
torques within clumpy models of cores and molecular clouds. Our
models were based upon a highly inhomogeneous simulation of compressible
magnetohydrodynamic turbulence. A Reverse Monte-Carlo radiative transfer method was used to calculate
the the intensity and anisotropy of the internal radiation
field, and the subsequent grain alignment was computed for a power-law distribution sizes 
using the DDSCAT package for radiative
torques. The intensity and anisotropy of the intracloud radiation field
show large variations throughout the models, but are generally sufficient
to drive widespread grain alignment. The $P-I$ relations for our
models reproduce those seen in observations. We show that the degree
of polarization observed is extremely sensitive to the upper grain
size cut-off, and is less sensitive to changes
in the radiative anisotropy. Furthermore, despite a variety of dust
temperatures along a single line of sight through our core and amongst
dust grains of different sizes, the assumption of isothermality amongst
the aligned grains does not introduce a significant error. Our calculations
indicate that sub-mm polarization vectors can be reasonably good
tracers for the underlying magnetic field structure, even for relatively
dense clouds ($\av\sim10$ to the cloud center).

The current predictive power of the grain alignment theory should motivate future polarization observations
using the next generation of multi-wavelength sub-mm polarimeters
such as those proposed for SOFIA.
\end{abstract}


\section{INTRODUCTION\label{sec:INTRODUCTION}}

The role of magnetic fields in star formation has been the subject of
heated debate (see McKee \& Tan 2002, Ballesteros-Peredes et al. 2006).
Polarization arising from aligned grains provides the most convenient way
to study magnetic field topology within the clouds and cores and therefore
helps to answer the questions facing the theorists. 

Observations of polarized sub-mm emission from molecular 
clouds (Hildebrand et al. 2000
and references therein) are widely interpreted as emission from dust
grains aligned with the local magnetic field. Advances in sub-mm technology
have ensured that measurable degrees of polarization are seen in many
interstellar clouds and cores (detections at the $0.5-8$\% level are
typical, Dotson et al. 2006), as well as in a wide variety of other objects, such as
circumstellar disks and the envelopes of young stellar objects. However, sub-mm
polarimetry can only diagnose the magnetic field projected in the
plane of the sky, and must be combined with other techniques to extract
the full 3D magnetic field structure (see Yan \& Lazarian 2006).
Nevertheless, the ubiquity of both interstellar dust and magnetic
fields ensures that the burgeoning field of sub-mm polarimetry is
a profitable venture. 

Substantial advancements in the grain alignment theory (see Lazarian
2003, for a review) have made it possible to predict the degrees of
polarization and therefore compare theoretical and numerical models with
observations. While simple models like those in Cho \& Lazarian (2005, hereafter CL05) show
a good correspondence between observations and theory, it is time to perform more
sophisticated modeling to develop a more complete understanding of both dust 
grains and interstellar magnetic fields. This motivates our work below.

The challenge of understanding the theory behind
polarization from aligned grains has been a significant one and
much theoretical work has been done over the decades since the discovery
of polarized starlight (Hall 1949, Hiltner 1949). Great minds like Edward
Purcell and Lyman Spitzer worked productively in the field. Eventually,
the original paradigm based on the paramagnetic alignment of grains (Davis \& Greenstein 1967, Jones \& Spitzer 1967, Purcell 1979, Spitzer \& McGlynn 1979)
was changed to one in which radiative torques played a major role. The
picture that emerged seemed natural and self-consistent. 

The alignment by radiative torques was first discussed in the 1970s. Dolginov \& Mytrophanov (1976) proposed radiative
torques as a mechanism that should naturally arise from the differential 
extinction of starlight (see also an earlier work by Dolginov 1972). However,
it took 20 years for the mechanim to get its due recognition. This finally happened after
Bruce Draine modified his open source DDSCAT code (Draine \& Flatau 1994) 
to include the effect of radiative torques. In Draine \& Weingartner
(1996, 1997) the efficiency of radiative torques was proven for grains
of rather arbitrary irregular shapes. However, alignment by
the paramagnetic mechanism was shown to be frustrated by the
 tendency of grains to flip (Lazarian \& Draine 1999a,b). 

Not only is the radiative torque mechanism attractive on theoretical
grounds but it is also motivated by observations. For example, at
the interface between diffuse and dense material $(0<\av<3)$ in the
Taurus Dark Clouds, Whittet et al. (2001) found that despite a uniform
grain population, the peak near-infrared polarization wavelength correlated
well with extinction. Lazarian (2003) explained this in terms of the natural
extinction dependence of the radiative torque mechanism, preferentially
aligning larger grains (which contribute to polarization at longer
wavelengths) with increasing $\av$. 

In this paper we deal with the grain alignment in starless cores and
in molecular clouds without bright OB stars. This is the case that 
challenges the theory to the utmost degree (see discussion in
Lazarian, Goodman \& Myers 1997). The original theoretical predictions
for the radiative torques restricted the domain of radiative torques
to $A_V<3$, which made the observations  of sub-mm polarization from
cores with $A_V>10$ (Ward-Thompson et al. 2000, 2002) very  surprising.
However, CL05 showed that the efficiency of
radiative torques \textit{increases rapidly with grain size such that even a diluted and reddened intercloud radiation field can lead to grain alignment}. As it is well known
that grains get bigger in molecular clouds, CL05 managed to address the
aforementioned observational challenge. 

In fact, CL05
 performed the first detailed
grain alignment calculation for a starless core model that included
an accurate treatment of grain alignment via radiative torques. In
lieu of a detailed grain alignment model, previous investigations
have generally made \emph{ad hoc} assumptions about the alignment
of grains. For example, it has been assumed that all grains
are aligned (Heitsch 2001), or in the case of Padoan et al. (2001)
that there exists a critical extinction $\av\sim3$ beyond which grain
alignment is simply turned off. Without imposing these assumptions
CL05 found that, provided there are large grains present, a significant
degree of polarization may be detected from dark cores with extinctions
as large as $\av=10$. Prior to modelling the polarization from particular
objects CL05 determined the criterion for alignment, namely that the
regular rotation rate arising from a balance between gaseous collisions
and radiative torques must exceed the thermal tumbling rate of the
grain. The ratio of these rates depends very sensitively upon grain
size: in general there is a minimum grain radius for alignment, $a_{alg}$.
For grains to be aligned they must be larger than $a_{alg}$. This
requirement becomes increasingly difficult to meet when the radiation
field is significantly extincted, since this reduces the radiative
torques and implies a larger $a_{alg}.$ With extinction comes reddening,
which further diminishes the radiative torques acting on small grains,
since grains couple with the radiation field most effectively when
$2\pi a\sim\lambda$, where $\lambda$ is the wavelength. Reddening increases
the wavelengths at which most of the radiation is present and thus
increases $a_{alg}$. Also, increasing the density leads to an increase
of the collision rate between grains and gas, which further increases
$a_{alg}$ since larger grains are less susceptible to these collisional
events. CL05 gave an empirical formula which
relates the minimum aligned grain size $a_{alg}$ to the density $\nh$
and extinction $\av$,

\begin{equation}
a_{alg}=(\log\nh)^{3}(\av+5)/2800,\label{eq:cl05}
\end{equation}\\
where $a_{alg}$ is in microns. Strictly speaking, this formula is
only valid for the smooth, spherical clouds they considered, however
it highlights the fact that increasing both the density and extinction
inhibits grain alignment by increasing $a_{alg}$. 

Despite the successes of their approach, CL05 lacked a detailed treatment
of the radiative transfer, which (1) drives grain alignment, and (2)
heats the dust grains, which subsequently reradiate the energy in
the sub-mm. CL05 assumed a smooth 1D radially stratified core, adopting
the radiation field of Mathis, Mezger \& Panagia (1983) and a constant
degree of anisotropy. They were able to show that the large grains
hypothesized to exist in dense clouds are sufficiently sensitive to
highly reddened, extincted starlight to align with the magnetic field
throughout their core. While the results are highly illustrative,
one is left to wonder what role inhomogeneities play in \emph{clumpy}
prestellar cores. In what follows the term `clumpy' shall be understood
to mean structures with enhanced densities. These structures can loosely
be categorized as sheets, filaments and cores, although we make no
such distinction in this paper. As shown in Bethell et al. (2004)
the radiation fields in a variety of clumpy clouds differ in a number
of important respects from that in a uniform cloud. A few optically
thin lines of sight allow considerable amounts of radiation to stream
to great depths in clumpy objects with mean center-to-edge extinctions
of $\av=10$. At these fiducial extinctions the radiation field is
not only enhanced in intensity by orders of magnitude relative to
the uniform case, but it is also considerably bluer. It is worth noting
that introducing clumpiness leads to the possibility of self-shielding,
sometimes rendering the \emph{mass} less well illuminated than the
\emph{volume}, especially near the surface of the cloud. The principal purpose of this paper is therefore to
extend the work of CL05 to the cases of clumpy clouds, treating the
radiative transfer, dust heating and grain alignment in a detailed
self-consistent manner using numerical methods. 

This paper complements CL05 in the sense that combined they offer
two extreme cases of cloud structure. We imagine that the results
they yield should at least bracket those of real clouds. In order
to uderstand what factors determine the observed polarization signal,
we perform parameter studies. These involve varying the upper grain
size limit $a_{max}$, the degree of anisotropy of the radiation field
$\gamma$, and the grain temperatures $\tg.$ It is often desirable
to estimate $\gamma$ and $\tg$ rather than calculate them in detail,
and we explore what effects such simplifying assumptions have on the
results. In general we find that the detailed properties of the radiation
field inside our models are sufficient to produce widespread grain
alignment. Despite the inhomogeneous conditions along any one line
of sight through the core, we find that the observed polarization
vectors are potentially useful tracers of the projected mass-averaged magnetic field
over the range of column densities present in our clouds. Our polarization
results should offer further encouragement to those exploring prestellar cores
through polarized sub-mm emissions.

The structure of our paper is as follows. In \S \ref{sec:THE-MODEL-STARLESS}
we describe our clumpy models, which are meant to represent a prestellar
core and a larger molecular cloud. We also outline our choice of dust
and interstellar radiation field. In \S \ref{sec:RADIATIVE-TRANSFER}
we give a brief description of the method we use to solve the radiative
transfer equation in the turbulent core, and discuss the results relevant
to grain heating and alignment via radiative torques. Section \ref{sec:GRAIN-ALIGNMENT}
deals with the alignment criterion for grains and their emission of
polarized sub-mm radiation. Here we also discuss the basic polarization
results for both our core and molecular cloud models in terms of polarization
degree \emph{versus} sub-mm intensity plots, the so-called `$P-I$
relation'. We explore how our results are influenced by dust temperature,
radiative anisotropies, and the limits imposed on the grain-size distribution.
In \S \ref{sec:Spectrum-and-alignment} the polarization spectra
for the models are shown, and we discuss how well polarization vectors
align with the underlying projected magnetic field vectors. We discuss
the implications of our results in \S \ref{sec:Discussion}.

\section{THE CLUMPY MODELS\label{sec:THE-MODEL-STARLESS}}

In this section we describe how we construct models of a prestellar
core and a starless molecular cloud from a turbulent MHD simulation
meant to represent a turbulently driven region of the ISM. While
the limitations of the numerical simulations in terms of
representing the actual interstellar turbulence, in particularly,
in terms of Reynolds and magnetic Reynolds numbers are well known (see
Lazarian \& Cho 2004), 
this does not compromise the goals of our study. Below we are not
interested in the details of
injection and dissipation of energy, but in
finding to what extent the sub-mm polarimetry represents magnetic fields.
For this purpose we concentrate on studying the effect of clumpiness on
the observed polarization, which is the effect that is missing in the smooth
CL05 model.

To simplify
what follows, the density structures of both our clumpy core and molecular
cloud models are based on a single three-dimensional simulation of
turbulently driven MHD turbulence, described on a cubical grid with
dimensions $256\times256\times256$ cells. The simulation is described
in detail in Vazquez-Semadeni et al. (2005). Here we briefly mention
numerical methods and parameters. We use a total variation diminishing
scheme (Kim et al. 1999) for solving the isothermal MHD equations
in a periodic computational box. The Poisson equation for the self-gravity
of gas is solved by using the Fast Fourier Transform method. We drive
a turbulent flow with the half scale of the size of one computational
dimension, by adding velocity perturbations at an equal time interval.
To achieve a high degree of clumpiness we adjust the velocity amplitude
in such a way that the root-mean-square sonic Mach number is equal
to around 10. The ratio of the initial magnetic pressure to gas pressure
is 0.1. The strength of the magnetic field is in a magnetically supercritical
range. The Jeans number, defined by the ratio of one dimensional size
of the computational box to the Jean length, is 4. 

We shall first discuss the construction of our core model. Although
our turbulent simulation possesses the desired (large) degree of clumpiness,
it is also statistically homogeneous and therefore not a good model
for a prestellar core.  As a rule, cores
obtained with direct numerical simulations are small and rather smooth, and
do not allow the studies that we have in mind. Therefore,
instead of using a brute force approach, we create a model core
combining numerical simulations and information obtained from observations.
  
\begin{figure}

\epsscale{0.7}

%\plotone{./figs/fig_mag_field_map.eps}
\plotone{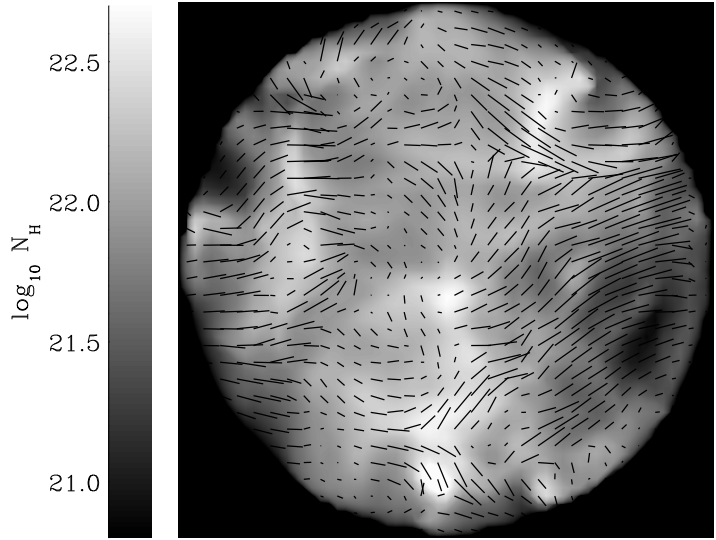}
\caption{ The column density $N_H$ of the model core (and molecular cloud).  The projected mass-weighted magnetic field vectors are shown superposed.\label{fig:denmap} }

\epsscale{1.0}

\end{figure}

Self-gravitating cores, although poorly resolved
in observations, exhibit central density peaks and lower density envelopes.
To improve our model we use the statistically homogeneous yet clumpy
density structure and follow CL05 by imposing both a spherical outer
boundary and a radial envelope profile. The underlying turbulent density
continuum is multiplied by the radial envelope function to endow the
cloud with a \emph{mean} radial profile%
\footnote{The mean radial profile is the average density in thin shells of radii
$r$ centered on the core center.%
} consistent with observations (Tafalla et al. 2004). The envelope
function is defined as;

\begin{displaymath}
\rho (r) \propto \left\{\begin{array}{ll}\mbox{const} & \mbox{if} \; r<r_0/4.7, \\
r^{-1} & \mbox{oterwise}.\end{array}
\right.
\end{displaymath}\\where the radius $R$ of the cloud is $R=24r_{0}=0.02$pc, and $r$
is measured from the center of the core. The spherical, radially stratified,
turbulent density continuum is then normalised such that the mean
visible band \emph{center-to-edge} extinction is $\av=10$, corresponding
to a column density of $N_{H}\sim1.7\times10^{22}$cm$^{-2}$ and
mean volume density $<\nh>\sim1.3\times10^{5}$cm$^{-3}$. Both the
physical size and mean column density are consistent with the cores
L183, L1544 and L43 observed in Crutcher et al. (2004). The column
density map of the core and its projected magnetic field are shown
in Figure \ref{fig:denmap}. The projected mass-weighted magnetic
field structure has been smoothed by a beamwidth equivalent to $1/32$
of the map width. From the column density map it is clear that the
large density fluctuations in this particular MHD simulation overwhelm
the imposed radial density gradient. While low mass prestellar cores
of this sort do not exhibit the large supersonic linewidths present
in the underlying MHD simulation, there is evidence that they are
clumpy nevertheless. The turbulent structure in this case
can be considered a proxy for clumpiness arising from the evolution
of localised gravitational collapse\footnote{An alternative approach to 
defining
a core would be to extract a core from a self-gravitating MHD simulation.
The MHD simulation used in this paper is not sufficiently evolved
to produce a large number of such cores, and the cores that are produced
are not well resolved numerically. For the pruposes of calculating
the internal radiation field it is important to resolve all structures
down to scales corresponding to an optical depth of order unity. For
this reason we use the entire MHD simulation to construct
our core.}.

When the opacity simply scales with density, as it does in our case,
the radiative transfer equation becomes a function of column density
only. Provided we preserve the column density structure, the internal
radiation field and any related quantities which do not depend explicitly
on density can then be used for a large family of objects; from small
dense cores to larger, more diffuse molecular clouds. Exploiting this
fact we rescale our model core to larger physical dimensions and reciprocally
decrease its density, such that the column density is held constant,
$N_{H}=<\nh>R=constant.$ Thus, from our model core we create a complementary molecular
cloud model with mean density $<\nh>=10^{3}$cm$^{-3}$ and physical
size $R=2.6$pc. The column density map and magnetic field structure
of the molecular cloud model are identical to that shown for the core
in Figure \ref{fig:denmap}. Not only will the internal mean intensity
and its anisotropy be the same for both models, but the resulting
temperatures of comparable dust grains (if present) will also be the
same. In order to exploit this invariance between core and molecular
cloud we have implicitly assumed that the extinction curves underlying
the radiative transfer through the two models are identical. This
is unlikely to be exactly true since dust evolves according to its environment
(Mathis 1990), the primary effect of which is observed as a change
in the ratio of total to selective extinction, $R_{V}\equiv\av/E(B-V)$.
In dense objects the small grains tend to grow through coagulation,
resulting in a flatter extinction curve and a larger $R_{V}.$ This
inconsistency may not prove to be problematic since the radiation
field in clumpy objects is determined by radiation streaming through
a optically thin windows. This radiation typically carries a relatively
weak imprint of the extinction curve; it is the geometry rather than
extinction curve which dominates the transfer of radiation. For comparably
clumpy clouds, Bethell et al. (2006) have shown that varying $R_{V}$
from $3.1$ to $5.5$ changes the intracloud V-band radiation field
by less than a factor of 2. This is in stark contrast to the case of uniform clouds,
where the internal radiation field is highly sensitive to the adopted
value of $R_{V}$ (Cecchi-Pestellini et al 1995).

\subsection{Dust properties\label{sub:Dust-Properties} }

To calculate as self-consistently as possible the attenuation of the
interstellar radiation field by dust, the subsequent equilibrium dust
temperatures, and the sub-mm emission, we use a detailed dust grain
ensemble composed of two materials; pure `astronomical' silicate and
pure graphite (Draine \& Lee 1984)%
\footnote{Tabulated optical properties may be found at the website www.astro.princeton.edu/\textasciitilde{}draine.%
}. In order to form a grain population it is necessary to apply a size
distribution to the individual grains; we select the simple power-law
grain size distribution of Mathis, Rumpl \& Norsdieck (1977, hereafter
MRN),

\begin{equation}
n(a)=A_{g,s}a^{-3.5}\label{eq:mrn}
\end{equation}\\
where $a$ is the grain radius, and $n(a)$d$a$ the number density
of grains with radii in the interval $[a,a+$d$a]$. The $A_{g,s}$
are the MRN abundance factors for silicates (\emph{s}) and graphite
(\emph{g}). Our adopted default lower and upper grain size cutoffs
(for both graphite and silicate) are $a_{min}=0.01\mu$m and $a_{max}=2\mu$m respectively for the
model core, and $a_{min}=0.005\mu$m and $a_{max}=0.5\mu$m for the
molecular cloud model. In order to adjust the limits of the grain
size distribution one must simultaneously ensure that the total dust
mass is conserved, which one can do by simply rescaling $A_{g,s}$.
In the MRN distribution the total number of dust grains is determined
by $a_{min},$ while the total mass in dust $M_{dust}$ is determined
largely by $a_{max}$, provided $a_{max}>>a_{min}$;

\begin{equation}
M_{dust}\propto\int_{a_{min}}^{a_{max}}A\, n(a)a^{3}da\propto Aa_{max}^{0.5}.\label{eq:dust_mass}\end{equation}\\
In order to avoid violating the total metal abundance we conserve
$M_{dust}$ by requiring that $A\propto a_{max}^{-0.5}.$ This point
is largely irrelevant to this paper, since we concern ourselves primarily
with \emph{relative} abundances of small and large grains, rather
than their total abundances.

The resulting grain populations produce extinction curves with total-to-selective
extinction ratios of $R_{V}\sim4$ and $5.5$ for the model molecular
cloud and core respectively. The upper grain size limit $a_{max}$
is rather larger than that presented in the original MRN. In the dense
ISM, dust grains are expected to increase their size through coagulation
(Clayton \& Mathis 1988, Vrba et al. 1993) and mantle growth (e.g.
Barlow 1978). By considering only large grains we restrict ourselves
to the sub-mm realm of dust emission. Shortward of $\lambda\sim100\mu$m
dust emission is dominated by very small grains heated stochastically
to high temperatures by single ultraviolet (UV) photons (e.g. Draine
\& Li 2001). The emission in the spectral regime dominated by stochastic
grain-heating is not well modelled by grains in thermal equilibrium,
the approximation we use in this paper, and which is only valid for
the regime in which large grains dominate the sub-mm emission ($\lambda>100\mu$m).
However, such heating is essentially irrelevant for our models if
we assume a sufficient depletion of very small grains.

Once the dust ensemble is fully prescribed and the ambient radiation
field known, it is then possible to calculate the equilibrium grain
temperature and spectral emissivity of each grain type and grain size at each
location in the core. At sub-mm wavelengths the core is optically
thin and the effects of absorption and scattering on the emergent
sub-mm flux can be ignored.

\subsection{The interstellar radiation field\label{sub:The-interstellar-radiation}}

Since our models are starless, the only permeating radiation field
is that of the highly diluted starlight which pervades the Galaxy.
The interstellar radiation field impinging upon our core is that of
Mathis, Mezger \& Panagia (1983) and consists of a superposition of
three diluted blackbody spectra and a small ultraviolet feature. The
blackbody temperatures are $T=3000,\,4000$ and $7500$K, and their
respective dilution factors $W$=$4\times10^{-13},1\times10^{-13}$
and $1\times10^{-14}$. These values are approximately valid for the
ISRF located at a Galactic radius $r_{G}=5$kpc. Additionally, we adopt
the simplifying assumption that the field is homogeneous and isotropic
in the vicinity of our core. In reality a small degree of intrinisc
anisotropy $(\sim10$\%) and inhomogeneity is expected in the ISRF,
resulting from the complex transfer of starlight through the Galaxy
as a whole. To simplify the radiative transfer through our core we
ignore this small intrinisic anisotropy, considering it to be of secondary
importance to the effects of clumpiness.

\section{RADIATIVE TRANSFER\label{sec:RADIATIVE-TRANSFER}}

It is here that our method differs most significantly from those in
CL05. In CL05 both the impingent and internal radiation fields were simply taken from Mathis,
Mezger \& Panagia (1983), which was possible because in one dimensional
models the radiation field is a function of $\av$ only (defined inwards
from the cloud surface). In order to calculate the penetration of
our clouds by the ISRF we use the three dimensional reverse Monte
Carlo numerical method described in Bethell et al. (2004). The \emph{reverse}
Monte Carlo scheme is based upon the reciprocity principle (Case 1957)
and differs from the more conventional \emph{forwards} scheme (e.g.
Witt 1977) in that it grows probabilistic photon trajectories in a backward
fashion. We start a reverse trajectory at what is its final physical
destination $\xb$ and direction $\kb$, evolving it probabilistically
until it reaches the external source of photons (the ISRF). It is
therefore well suited to describing not only the mean intensity but
also the angular distribution of the intensity $I(\kb,\xb)$ at any
chosen point. The moments of the radiation field, and in particular
the degree of anisotropy (described later), are obtained straightforwardly
from $I(\kb,\xb)$. Trajectories are generated until $I(\kb,\xb)$
meets a prescribed signal-to-noise criterion (in this paper the Monte
Carlo noise levels are uniformly kept below 1\%). Algorithmic control
of the signal-to-noise ratio is of particular importance for dense,
self-shielding clumps where most of the mass resides, because these
may be relatively poorly explored in simple forwards Monte Carlo schemes.

The individual trajectories are constructed from contiguous line segments,
the length of each is a probabilistic sampling of the optical depths
due to pure scattering, while the angles between segments are drawn
probabilistically from the scattering redistribution function. The
redistribution function is taken to be the popular Henyey-Greenstein
function (Henyey \& Greenstein 1941). While not the most sophisticated
phase function (Witt 1977, Draine 2003) it encapsulates the essential
features of scattering, and takes a conveniently compact mathematical
form. It is described by one parameter, the scattering anisotropy
$g=<\cos\theta>$, which lies in the interval $[-1,1]$ where $g=-1$
represents backwards scattering, $g=0$ isotropic scattering, and
$g=1$ forwards scattering. While observations of reflection nebulae
and diffuse Galactic emission (Gordon 2004 and references therein)
place constraints on the dust albedo $\omega$ and anisotropy parameter
$g$, the important role of inhomogeneities tends to increase the
uncertainties in the values derived from radiative transfer modelling
(Mathis, Whitney \& Wood 2002). For the dust ensemble described in
\ref{sub:Dust-Properties} the values of $g$ are typically forward
throwing, ranging from 0.5 in the visual to 0.75 in the UV.

The quantity most relevant to grain heating is the specific mean intensity,
defined as;

\begin{equation}
\jlamx=\frac{{1}}{4\pi}\int\ilamkx\quad d\Omega,\label{eq:jlambda}\end{equation}\\
where $\Omega$ is the solid angle in steradians. The radiative equilibrium
temperature $\tg$ of an individual dust grain is determined by balancing
the absorbed radiation (UV-visible) with that emitted in the sub-mm
(see equation 6.2 in Draine \& Lee 1984). By integrating the spectral
emissivity $j_{\lambda}(a,T_{g},\xb)$ over the dust size distribution
we obtain the total spectral emissivity $j_{\lambda}(\xb)$ in ergs
cm$^{-3}$ sterad$^{-1}$ cm$^{-1}$. 

In order to compute the radiative torque acting on a dust grain we
first calculate the spectral degree of anisotropy of the radiation
field $\glamx$, given by;

\begin{equation}
\glamx=\frac{1}{4\pi\jlamx}|\int\ilamkx\kb d\Omega)|.\label{eq:gamma_lambda}\end{equation}\\
 The spectral degree of anisotropy $\glamx$ lies in the range $[0,1]$,
where 0 corresponds to an isotropic field and 1 to a unidirectional
field; in essence it measures the directionality of the radiation
field. We further simplify $\glamx$ by defining the wavelength independent
bolometric anisotropy;

\begin{equation}
\gx=\frac{{\int\glamx\jlamx d\lambda}}{\int\jlamx d\lambda},\label{eq:gamma_bol}\end{equation}\\
which also lies in the range $[0,1]$, and is the quantity used in
the radiative torque calculations in Section \ref{sec:GRAIN-ALIGNMENT}.
The UV-visible anisotropy in a clumpy medium depends largely on the
clumpy geometry, rather than the wavelength dependent properties of
dust scattering. In this case the anisotropy is approximately wavelength
independent and $\gamma_{\lambda}(\xb)\sim\gamma(\xb)$ for most wavelengths.

\subsection{Results\label{sub:Results_rt}}

\begin{figure}

%\plotone{./figs/fig_triplet_den_j_dip_maps.eps}
\plotone{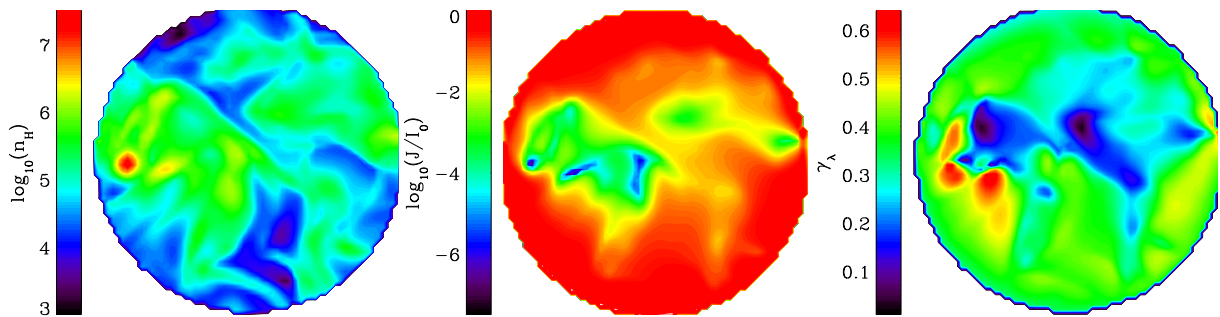}
\caption{\textit{Left}: a slice through the density continuum of our model core.  The corresponding slice through the molecular cloud model is identical except that the densities are a factor of 130 smaller. \textit{Center}: the same slice through the internal mean intensity $J_{\lambda}$ at $\lambda=0.54\mu$m, normalised to the value of the unattenuated ISRF, $I_{0}$. \textit{Right}: the anisotropy of the radiation field, $\gamma$, at 0.54$\mu$m. \label{fig:triplet_den_j_dip}}

\end{figure}

\begin{figure}

\epsscale{0.6}

%\plotone{./figs/fig_jv_distribution.eps}
\plotone{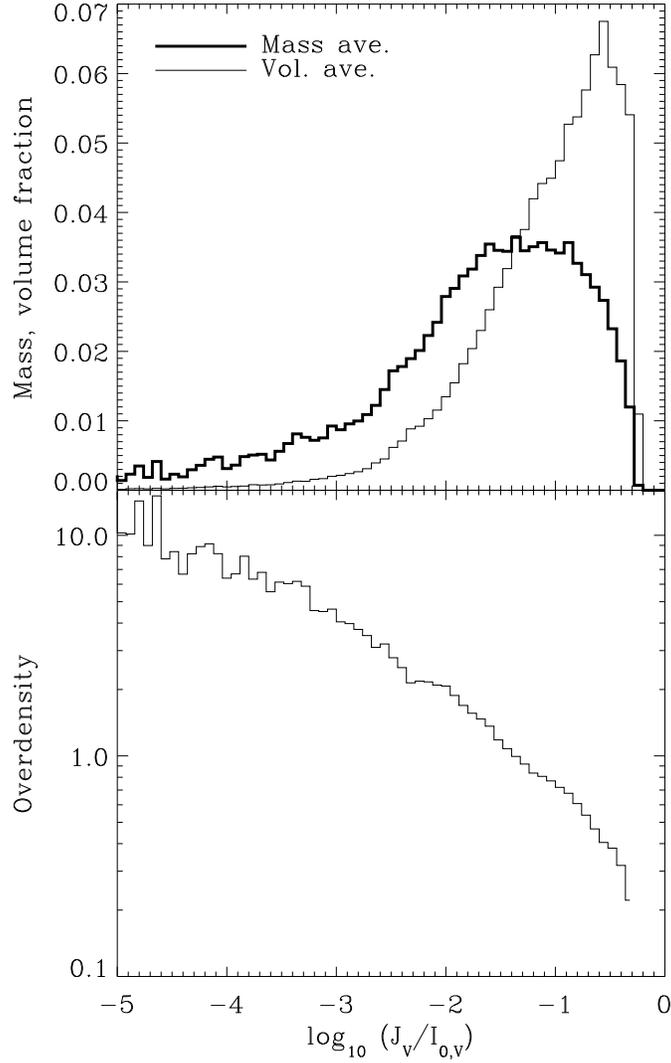}
\epsscale{1.0}

\caption{\textit{Top panel}: The fractions of the mass (\textit{thick line}) and volume (\textit{thin line}) associated with an internal V-band relative mean intensity $J_{V}/I_{0,V}$. The quantity $I_{0,V}$ is the unattenuated ISRF intensity. \textit{Bottom panel}: The mean overdensity (i.e. the mass fraction divided by the volume fraction) as a function of the mean intensity.  For example, the overdensity indicates that material bathed in radiation for which $\log_{10}(J_{V}/I_{0,V})=-4$ is, on average, seven times more dense than the global average density. Each bin is 0.08dex in width.\label{fig:jv_dist}}

\end{figure}

\begin{figure}

%\plotone{./figs/fig_dipole_distribution.eps}
\plotone{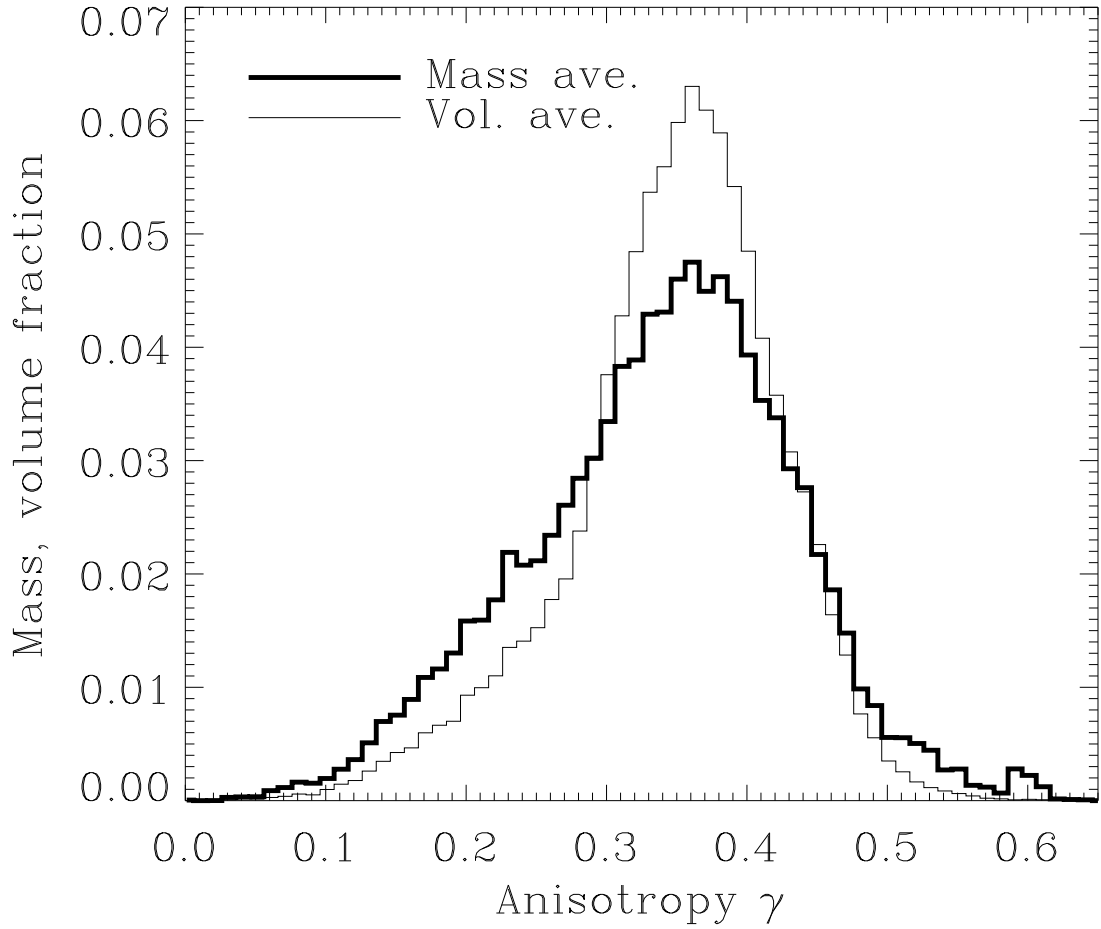}
\caption{The mass (\textit{thick line}) and volume (\textit{thin line}) fractions associated with a radiative anisotropy $\gamma.$  Both the mass and volume averaged anisotropies are approximately equal at $\gamma_{m,v}\sim0.34$. \label{fig:dip_dist}}

\end{figure}

Before proceeding with calculations of grain alignment and polarized
dust emission it is worth considering some of the underlying radiative
transfer results. In Figure \ref{fig:triplet_den_j_dip} we show comparative
slices of the density continuum $\nh(\xb)$, the internal mean intensity
$\jlamx$, and the anisotropy $\glamx$, evaluated at $\lambda=0.54\mu$m.
The morphology of the radiation field is of interest, as it indicates
which environments possess favorable conditions for grain alignment.
In a uniform cloud $\jlamx$ decreases in a quasi-exponential manner
with increasing depth from the cloud surface (Flannery et al. 1980).
In the clumpy cloud we generally see a much weaker variation with
radial position, and a more appreciable anticorrelation with density.
The radiation field floods the available volume, illuminating the
surfaces of clumps deep inside the cloud. Much of the attenuation
of the radiation field occurs in the envelopes of our self-shielding
clumps, where $\jlamx$ may drop precipitously by several orders of
magnitude. The heirarchical structure of our clumpy cloud ensures
that the densest points are the most shielded, and extremely large
extinctions can occur over small distances in the vicinity of clumpy
structures. The model mass and volume fractions as functions of $J_{V}$
are shown in the top panel of Figure \ref{fig:jv_dist}. In the lower
panel we show the mean overdensity of material associated with a value
of $J_{V}$, which illustrates clearly the effect of self-shielding.
The very darkest locations are on average about ten times denser than
the global mean density. Referring back to Figure \ref{fig:triplet_den_j_dip}
we see that the degree of anisotropy shows morphologically very little
correlation with density, and tends to be largest on the \emph{surfaces}
of sheets, filaments and clumps, where gradients in $\nh$ and $\jlamx$
are usually large. It is worth noting that the anisotropies found
in clumpy clouds are not necessarily larger than those found in uniform
clouds. In a uniform spherical cloud there is always a single well-defined
`brightest direction' (except at the cloud center): the anisotropy
$\gamma\sim0.5$ at the cloud surface, initially increasing slightly
with depth, only to decrease towards $\gamma=0$ at the cloud center.
In contrast, the anisotropy in the clumpy cloud shows little radial
dependence. Instead, the anistropy at a point is dominated by a few
beams of starlight distributed quasi-randomly in the sky. These fluxes
tend to average out somewhat, resulting in rather small mass-averaged
anisotropies of $\gamma\sim0.34$. To illustrate this point, we plot
the mass and volume fractions as a function of $\gamma$ in Figure
\ref{fig:dip_dist}. Both curves are peaked at $\gamma\sim0.3,$ although
the mass fraction curve is slightly broader. The similarity between
the mass and volume-weighted curves indicates that there is little
overall correlation between density and anisotropy. Furthermore, the
anisotropy varies by less than an order of magnitude, generally falling
within the range $\gamma=0.1-0.6$. These values are somewhat lower
than the constant value $\gamma=0.7$ adopted by CL05 and Draine \&
Weingartner (1996) for their dense clouds. In a later section we shall
explore how assuming a spatially uniform value for $\gamma$ affects
the polarized emission.

\begin{figure}

%\plotone{./figs/fig_tg_dist.eps}
\plotone{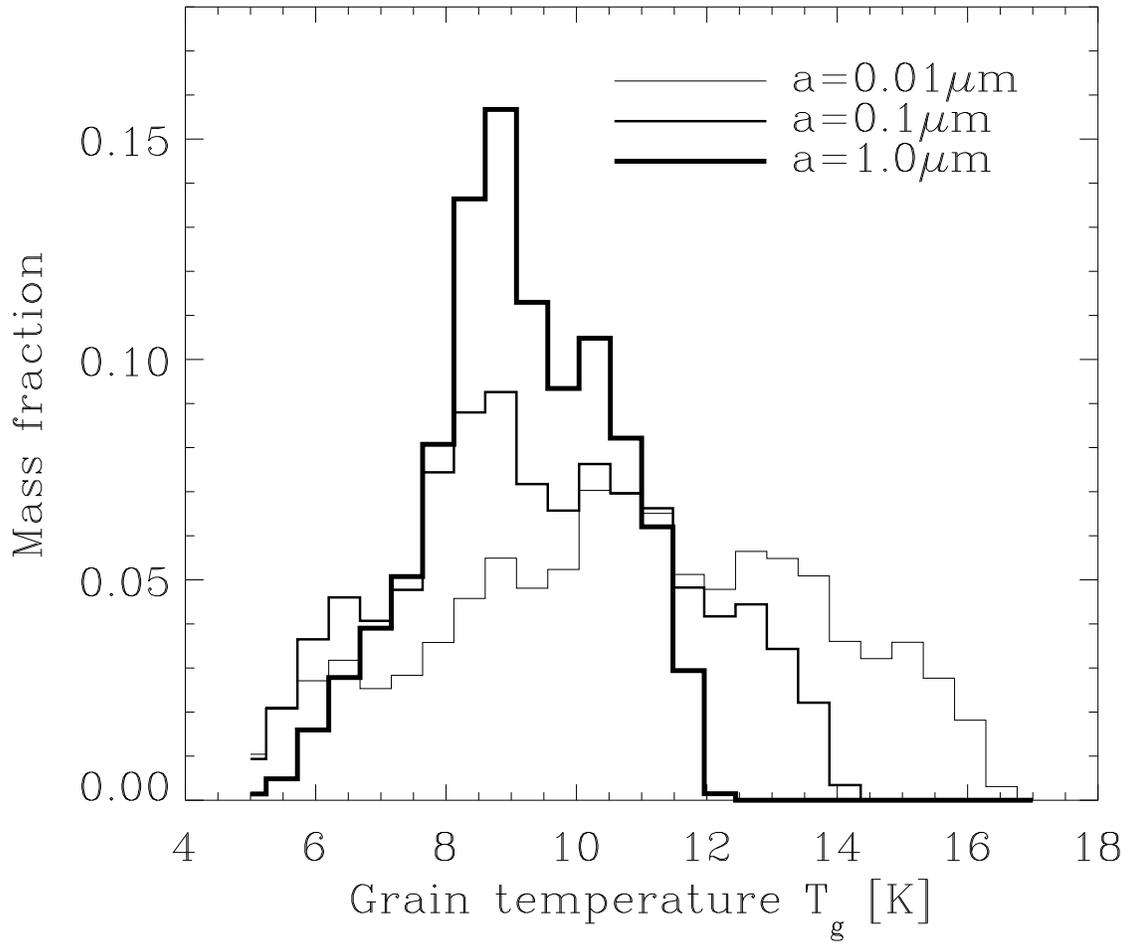}
\caption{The distribution of mass with silicate grain temperature for grains of radii $a=0.01,0.1$ and $1.0$ $\mu$m.  The binsize is 0.48K. \label{fig:tg_dist}}

\end{figure}

The temperature distributions of graphite grains with radii $a=0.01,\,0.1$
and $1.0\mu$m are shown in Figure \ref{fig:tg_dist}. The inefficiency
with which small grains emit sub-mm radiation accounts for the systematic
increase of grain temperature with decreasing grain radius. Furthermore,
small grains preferentially absorb the bluer part of the ISRF, which,
due to the steep increase of extinction with decreasing wavelength,
is subject to the largest spatial variations inside the core. The
relatively wide temperature distribution of these small grains reflects
this sensitivity. The assumption of isothermal dust temperatures is
therefore best applied when the emission is dominated by the large
grains (i.e. at $\lambda>300\mu$m). At shorter wavelengths the emission
is dominated by the warm, small grains, and the extraction of useful
physical properties of a cloud is complicated by the need to know
the temperature distribution of the grains (Hildebrand 1983, Chen
1990, Li et al. 1999, Schnee et al. 2006). In later sections we shall
explore the effect an isothermal dust temperature assumption has on
the polarization degree.

\section{GRAIN ALIGNMENT AND POLARIZED SUB-MM EMISSION\label{sec:GRAIN-ALIGNMENT}}

We are now in a position to begin calculating grain alignment and
the polarized emission from these grains. For the relatively large
grains we consider in this work, the damping of a constant radiative
torque leads to a grain spinning with a steady angular frequency $\omrad$.
However, in the absence of a strong radiative torque one would expect
collisions with gaseous atoms to cause the grain to rotate irregularly
with an angular velocity $\omt$, corresponding to a kinetic energy
$\sim kT/2$. Only when the grain is rotating suprathermally, i.e.
$\Omega_{rad}>\Omega_{th}$, will the damping from gaseous collisions
be insufficient to prevent the grain from aligning with the magnetic
field. Following CL05 we consider a grain to be aligned with the magnetic
field if its angular frequency of rotation is greater than the thermal
value by a factor of three, i.e. $\Omega_{rad}/\Omega_{th}>3$. The
ratio $\Omega_{rad}/\Omega_{th}$ is given by;

\begin{equation}
(\Omega_{rad}/\Omega_{th})^{2}=\frac{5\alpha_{1}}{192\delta_{2}}(\frac{1}{\nh kT})^{2}\frac{\rho a_{eff}}{m_{h}}[\int d\lambda Q_{\Gamma}\lambda(4\pi\gamma J_{\lambda}/c)]^{2}(\frac{{1}}{1+\tau_{d,gas}/\tau_{d,em}})^{2}\label{eq:criterion}\end{equation}\\
where $a_{eff}$ is the effective grain cross-section, and $\tau_{d,gas}/\tau_{d,em}$
is the ratio of gas drag to thermal emission drag times. The quantities
$c$ and $k$ are the speed of light and Boltzmann's constant respectively.
The radiative torque efficiency, $Q_{\Gamma},$ is calculated using
the discrete dipole code DDSCAT (Draine \& Flatau 2004). The anisotropy
and mean intensity appear in the integral as the product $\gamma J_{\lambda},$
and this along with $Q_{\Gamma}$ determines the strength of the radiative
torque. The smallest grain size for which the above criterion is met
is denoted by $a_{alg}.$ Typically, grains of radii larger than $a_{alg}$
will be aligned with the magnetic field, while smaller grains will
not. When a grain is considered aligned we assume its alignment with
the magnetic field is perfect.

Having defined the condition for grain alignment, we proceed to calculate
the polarization signal arising from the thermal emission of the spinning
grains. At sub-mm wavelengths the core is optically thin to the emitted
radiation from dust, and to a good approximation the emergent sub-mm
intensity is simply the integral of the dust emissivity along a lines
of sight through the cloud. Following Fiege \& Pudritz (2000), we
write out the Stokes parameters of the emitted radiation as functions
of plane-of-sky coordinates $(\psi,\phi)$: 

\begin{equation}
Q=\int\nh\, G_{pol}\cos2\psi\cos^{2}\phi dz,\label{eq:Q}\end{equation}

\begin{equation}
U=\int\nh\, G_{pol}\sin2\psi\cos^{2}\phi dz,\label{eq:U}\end{equation}

and

\begin{equation}
I=\int\nh\, G_{ran}dz-\int\nh\, G_{pol}(\frac{1}{2}cos^{2}\phi-\frac{1}{3})dz.\label{eq:I}\end{equation}\\
The density $\nh$ is the appropriate weighting for the dust emission
along the line of sight, since the dust-to-gas ratio is assumed to
be constant. The quantity $\psi$ is the angle between the plane-of-sky
projection of the magnetic field and north, and $\phi$ is the angle
between the magnetic field and the picture plane. The polarized and
unpolarized components of the emissivity are respectively

\begin{equation}
G_{pol}=\sum_{i}\sigma_{pol,i}\int_{a_{align,i}}^{\infty}a^{3}B_{\lambda}(T_{g})n_{i}(a)da\label{eq:Gpol}\end{equation}\\
and
\begin{equation}
G_{ran}=\sum_{i}\sigma_{ran,i}\int_{0}^{\infty}a^{3}B_{\lambda}(T_{g})n_{i}(a)da,\label{eq:Gran}\end{equation}\\
where $B_{\lambda}(\lambda)$ isthe blackbody function. As described
in Section \ref{sub:Dust-Properties} $n_{i}(a)$ is the grain number
density. The following approximation is used in eqns \ref{eq:Gpol}
and \ref{eq:Gran} to simplify the sub-mm emission efficiencies of
the grains; 

\begin{equation}
C_{X,i}=\sigma_{X,i}\lambda^{-2}a^{3},\quad a<<\lambda\label{eq:cross-sec}\end{equation}\\
where $X$ denotes either the polarized ($pol$) or random ($ran$)
components, and the index $i$ denotes silicate ($s$) or graphite
($g$). For this calculation the relative abundances of the polarizing
and non-polarizing cross-sections (Lee \& Draine 1984) are related
via $\sigma_{pol,s}/\sigma_{ran,s}=0.67,\,\sigma_{ran,g}/\sigma_{ran,s}=1.60,$
and $\sigma_{pol,g}=0.$ Here we have assumed that the graphite grains
do not align with the magnetic field. The degree of polarization $P$
follows from the Stokes vector,

\begin{equation}
P=\frac{\sqrt{Q^{2}+U^{2}}}{I}.\label{eq:P}\end{equation}

Wherever alignment prevails we assume it is perfect, and the angle
$\theta$ between the grain angular momentum vector and the magnetic
field is zero. The Rayleigh reduction factor, $R=1.5(<\cos^{2}\theta>-1/3),$
which measures the effective degree of alignment between the population
of grains and magnetic field, is simply reinterpreted as the fraction
of the total emissive dust cross-section associated with aligned grains;

\begin{equation}
R=\frac{{\int_{a_{alg}}^{a_{max}}C_{ran}}n(a)\, da}{\int_{a_{min}}^{a_{max}}C_{ran}n(a)\, da}.\label{eq:R}\end{equation}\\
By increasing $a_{max}$ we increase this fraction. The Rayleigh reduction
factor is now a function of position on our core; $a_{alg}$ is the
smallest grain which satisfies the alignment criterion $\Omega_{rad}/\Omega_{th}>3$,
and so depends on the local radiation field and density.

\subsection{Results for the model core\label{sub:Results:}}

\begin{figure}

\epsscale{0.7}

%\plotone{./figs/fig_850_map_pdeg.eps}
\plotone{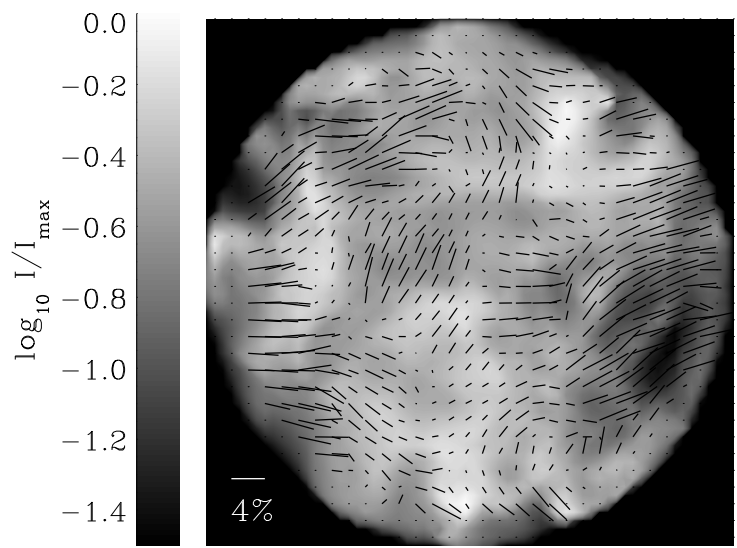}
\caption{The 850$\mu$m emission map of the model core.  Superposed are the projected polarization degree vectors. A $4\%$ polarization vector is shown for scale.\label{fig:850_map}}

\epsscale{0.7}

\end{figure}

Our default core model has the density structure described in Section
\ref{sec:THE-MODEL-STARLESS}, and an MRN grain size distribution
with $a_{max}=2\mu$m. The anisotropy and mean intensity of the internal
radiation field (see Section \ref{sec:RADIATIVE-TRANSFER}) is used
to determine the alignment and polarized emission from grains throughout
the model. Unless otherwise stated the results are made at an observational
wavelength of $850\mu$m. The emission map of our default core model,
including polarization vectors, is shown in Figure \ref{fig:850_map}.
The polarization vectors have been smoothed in a similar manner to
the projected magnetic field, i.e. by a beamwidth equal to $1/32$
of the map width. The differences between the $850\mu$m map and those
computed at other wavelengths of interest, such as $350$ and $450\mu$m,
are visually very minor. In what follows we consider $850\mu$m our
default wavelength. Comparing the $850\mu$m map with the column density
$N_{H}$ and projected magnetic field (see Figure \ref{fig:denmap})
we see that the sub-mm emission map correlates well with the column
density. This can be explained by noting that $850\mu$m is in the
Rayleigh-Jeans spectral regime, combined with the fact that the large
grains responsible for the sub-mm emission possess a narrow distribution
in temperature (see Figure \ref{fig:tg_dist}). Under these circumstances
the volumetric emissivity (in units of ergs s$^{-1}$ cm$^{-3}$ $\mu$m$^{-1}$)
is almost directly proportional to the density, $\nh$. 

\begin{figure}

\epsscale{0.8}

%\plotone{./figs/fig_forcealign.eps}
\plotone{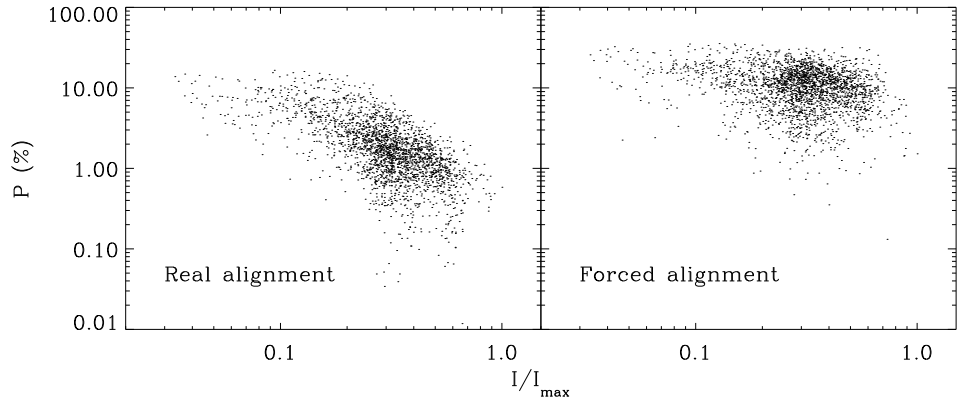}
\caption{\textit{Left}): The polarization degree, $P$, as a function of normalized intensity, $I/I_{0}$, for our core model.  This calculation uses the full, detailed internal radiation field and dust temperatures. \textit{Right}: The P-I relation obtained when all silicate grains are forced into alignment with the magnetic field.  Each dot represents a line of sight through the cloud, or equivalently a pixel in Figure \ref{fig:850_map}. These polarization degrees are evaluated at the default wavelength of 850 $\mu$m. \label{fig:poldeg_force_alignment}}

\epsscale{1.0}

\end{figure}

Sub-mm observations often reveal various {}``depolarization'' effects,
especially towards the centers of strong sub-mm emission (Gon\c{c}alves,
Galli \& Walmsley 2005). This appears as an empirical anticorrelation
between the polarization degree $P$ and the sub-mm intensity $I$,
and which is usually fitted with a power-law $P\propto I^{-\alpha}$
(Henning et al. 2001, Lai et al. 2003, Crutcher et al. 2005). The
index $\alpha$ is seen to vary considerably between cores, typically
taking values in the range 0.6-1.2 (Matthews \& Wilson 2000, 2002,
Henning et al. 2001, Lai et al. 2003, Crutcher et al. 2004). In the
case of $\alpha>1$ the polarized intensity $I_{p}=PI$ \emph{decreases}
towards bright sub-mm regions. The $P-I$ scatter for our core is
shown in the lefthand panel of Figure \ref{fig:poldeg_force_alignment}.
Qualitatively it exhibits the observed $P-I$ anticorrelation, although
there is also a large scatter about this trend. Similar plots were
produced in CL05 showing a tighter correlation, primarily due to the
smoothness of their cores. On the other hand the periphery of our
core exhibits so-called {}``polarization limb brightening'', in
which the conditions for alignment are favorable (i.e. a strong radiation
field and appreciable anisotropy) and the line of sight passes through
only a short section of material. These points represent the upper
limit in the $P-I$ relation at small $I/I_{max}.$ 

CL05 showed that the power-law index $\alpha$ is strongly dependent
on the upper cutoff in the grain size distribution $a_{max}$. Considering
that the largest grains are typically the first to be aligned, increasing
$a_{max}$ above the threshold for alignment $a_{alg}$ increases
the fraction of the grain population for which the alignment criterion
(equation \ref{eq:criterion}) is met. In the dense, dark clumps into
which longer wavelength radiation penetrates, only the very largest
grains are expected to be aligned via radiative torques. Under these
circumstances increasing $a_{max}$ produces preferentially more polarization
in regions of strong sub-mm emission. This is observed as a flatter
$P-I$ relation, characterized by smaller values of $\alpha$. It is important to clarify exactly how much of this relation is due to the
efficiency of grain alignment via radiative torques, as opposed to
effects attributable to changes in the magnetic field orientation
along lines of sight. In other words, what is the maximum observable polarization that can be obtained form this core?  To address this question we recalculate the polarization map
at $850\mu$m \emph{with all silicate grains aligned}, the result
of which is shown in the right panel in Figure \ref{fig:poldeg_force_alignment}.
In this case the scatter and $P-I$ relation is due to geometrical
depolarization effects along each line of sight, and represents the
maximum possible observational polarization we can obtain from the
model.  In particular one should note that at large $I/I_{max}$ some of the reduction in polarization
must be caused by the magnetic field morphology changing along the lines
of sight.

\subsubsection{Effect of changing $a_{max},\,\gamma$ and $\tg$\label{sub:Influence-of-gamma}}

Central to this paper is the detailed calculation of the anisotropy
$\gx$ and dust grain temperatures $T_{g}(\xb)$. Before we explore
the sensitivity of our results to these quantities, we first consider
the effects of changing the upper grain-size limit $a_{max}.$ The
existence of sufficiently large grains is essential if the radiative
torque mechanism is to produce detectable degrees of polarization
($P>1\%$). The extinction properties of dust in the very dense ISM
suggest that the grain size distribution extends to larger $a_{max}$,
which is believed to be due to the process of grain coagulation. The
original MRN distribution which reproduces the average Galactic extinction
curve has a sharp cutoff at $a_{max}=0.25\mu$m, although more recent
distributions include an exponential tail of large grains extending
up to $a>1\mu m$ (Weingartner \& Draine 2001). Provided the grain
size distribution extends beyond the minimum aligned grain size $a_{alg}$,
i.e. $a_{max}>a_{alg}$, a fraction of the grains will be aligned
and we can expect a non-zero degree of polarization. 

\begin{figure}

\epsscale{0.9}

%\plotone{./figs/fig_gamma_amax.eps}
\plotone{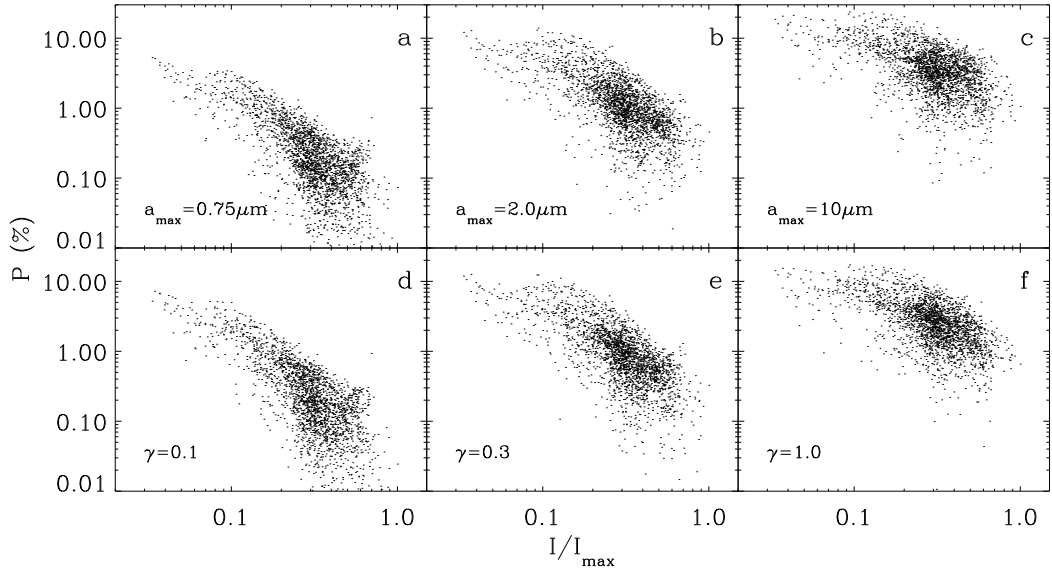}
\caption{P-I plots for the model core obtained from the parameter study. The top row (a-c) shows how varying the upper grain-size cutoff, $a_{max}$, changes the P-I relation. Plot $b$ is in fact the result for our default core model. In the bottom row (d-f) we have replaced the real, spatially varying values for the radiative anisotropy, $\gamma$, with constant values $\gamma=0.1,\,0.3$ and $1.0$.  For comparison, the true mass averaged anisotropy is $\gamma_m=0.34$.  These plots are evaluated at 850$\mu$m. \label{fig:poldeg_changing_dipole}}

\epsscale{1.0}

\end{figure}

The $P-I$ relations obtained using different values of $a_{max}$
are shown in Figure \ref{fig:poldeg_changing_dipole}. It is clear from this
figure that increasing $a_{max}$ increases the polarization degree.
Regions of high column density (i.e. the bright sub-mm regions) benefit
most from the increase in $a_{max}$ since these regions are expected
to be the least well illuminated by an anisotropic radiation field,
and an increase in $a_{max}$ under these conditions can make the
difference between having some grains aligned ($a_{max}>a_{alg}$)
and having no grains aligned ($a_{max}<a_{alg}$).

We can explore the effects of changing $\gx$ by recalculating the
polarization with constant values for the anisotropy, $\gamma=0.1,\:0.3$
and $1.0$, covering the range of real values (the real mass averaged
anisotropy is $\gamma_{m}\sim0.34$, see Figure \ref{fig:dip_dist}).
The resulting $P-I$ relations are shown in the bottom panel of Figure
\ref{fig:poldeg_changing_dipole}. Although changing $\gamma$ in this way
does little to the scatter in the $P-I$ relation it is clear that
it affects the overall scaling. For $\gamma=0.1$ (consistent with
the intrinsic anisotropy of the Galactic ISRF) the radiative torques
are suppressed to such an extent that few of the grains responsible
for the $850\mu$m emission are aligned and $P<1\%$. With $\gamma=0.3$
the results are quite similar to those obtained with the real $\gx$
values, and a significant fraction of the cloud has a degree of polarization
above $1\%
$. With the maximal anisotropy, $\gamma=1.0,$ we see a further increase
in $P$, the upper envelope of which is now above $1\%$ over the
entire range of observed $I/I_{max}.$ 

\begin{figure}

%plotone{./figs/fig_tempconst_histo_both.eps}
\plotone{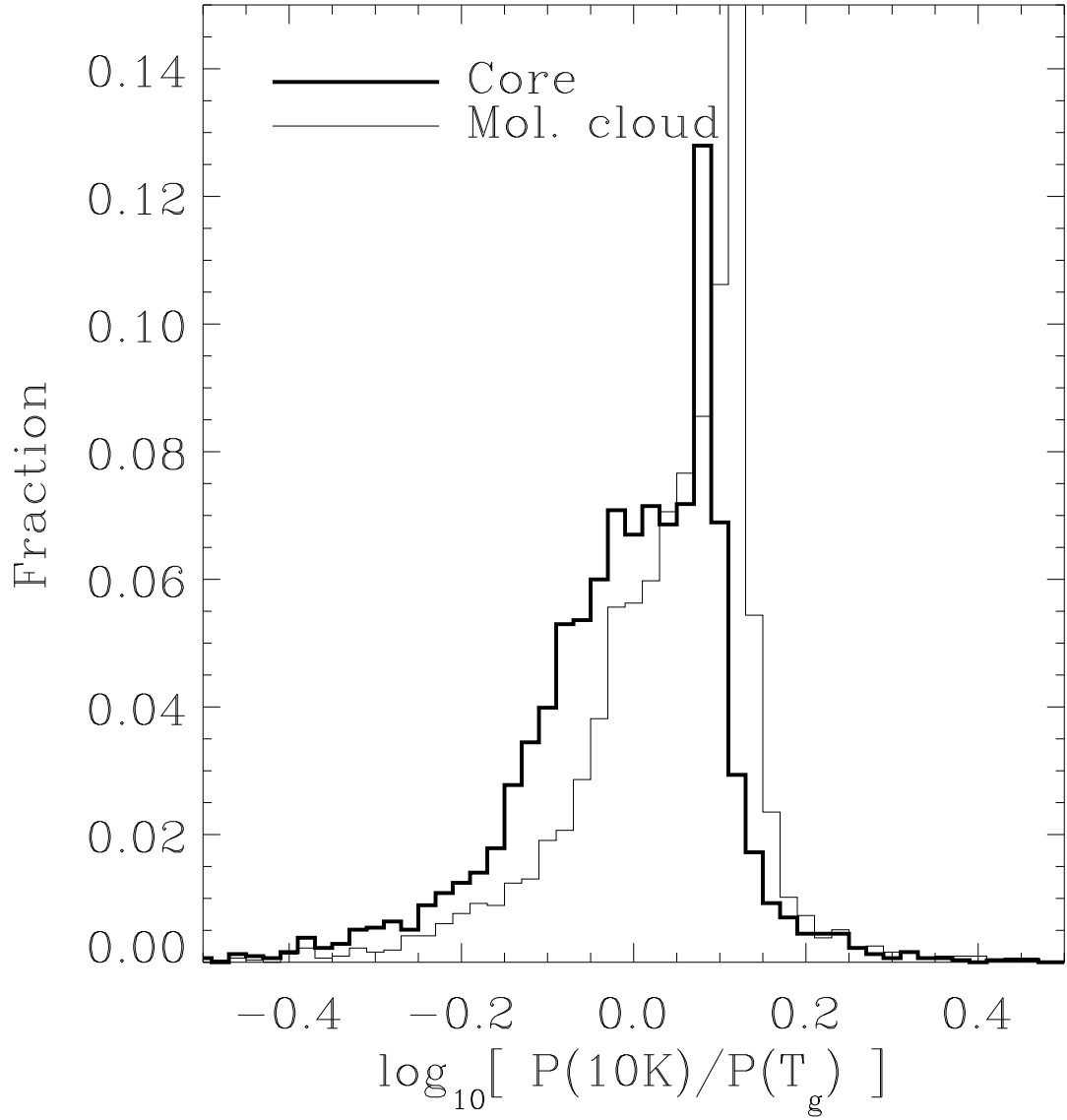}
\caption{The discrepancy in polarization degrees obtained using the true grain temperatures, $T_g$, and holding them fixed at 10K.  The binsize is 0.02dex.\label{fig:poldeg_tempconst_histo}}

\end{figure}

The narrow temperature distribution of large grains (see Figure \ref{fig:tg_dist})
suggests that temperature variations amongst the grains emitting most
of the $850\mu$m emission are small and relatively unimportant. Indeed,
if we recompute the polarization degree using $T=10$K for all grains
regardless of size or location, we obtain a result almost indistinguishable
from that using the physically derived grain temperatures $\tg(a,\xb).$
This discrepancy is shown in histogram form in Figure \ref{fig:poldeg_tempconst_histo}.
The `isothermal' approximation leads to small errors (generally no larger than 0.2dex) in the $850\mu$m polarization
signal.

\subsection{Results for the molecular cloud model}

\begin{figure}

\epsscale{0.8}

%plotone{./figs/fig_forcealign_molcloud.eps}
\plotone{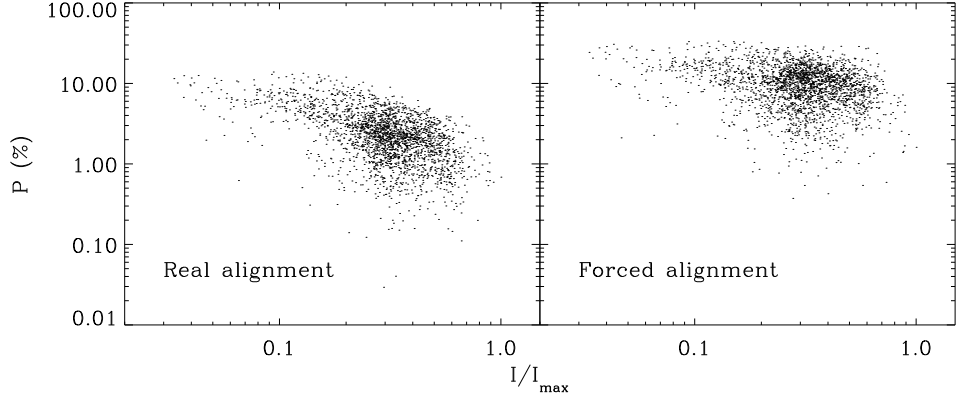}
\caption{\textit{Left}: The polarization degree, $P$, as a function of normalized intensity, $I/I_{0}$, for our molecular cloud model. \textit{Right}: The P-I relation obtained when all silicate grains are forced into alignment with the magnetic field (\textit{right}).  Both are evaluated at 850 $\mu$m.\label{fig:poldeg_force_alignment_mc}}

\epsscale{1.0}

\end{figure}

The main differences between the molecular cloud and core models are
the densities involved and the range of $a_{max}$ considered. Our
molecular cloud model has $a_{max}=0.5\mu$m and a mean density
of $\nh=10^{3}$cm$^{-3}$, and exploits the same radiation field and
dust temperatures used in our core model. The $P-I$ relation for
our molecular cloud model is shown in the top panel of Figure \ref{fig:poldeg_force_alignment_mc}.
In the lower panel we show the maximal polarization arising when all
silicate grains are aligned with the magnetic field. It is informative
to compare these results with those for the default core model in
Figure\ref{fig:poldeg_force_alignment}. The two $P-I$ relations
are qualitatively similar, although it seems that a slightly larger
degree of polarization is obained at large $I/I_{max}$ from our default
molecular cloud model. As described in Section \ref{sec:GRAIN-ALIGNMENT},
a reduction in $\nh$ reduces the disalignment arising from collisions
between grains and gas-phase particles. This favors grain alignment
by lowering $a_{alg}$ (see equation \ref{eq:cl05}). Although the
model molecular cloud has fewer large grains ($a_{max}=0.5\mu$m)
it seems that the concomitant reduction in $a_{alg}$ is more than
enough to compensate. Applying equation \ref{eq:cl05} we see that
reducing the density by a factor of 130 (which is required in order
to turn our core into the molecular cloud) leads to a factor of $\sim10$
decrease in $a_{alg}$. This is larger than the factor of 4 decrease
in $a_{max}$ (from $2\mu$m to 0.5$\mu$m). As a result a larger
fraction of the grains in the molecular cloud are aligned. We now
perform a brief parameter study to see how sensitive these results
are to our default values.

\subsubsection{Effect of changing $a_{max},\,\gamma$ and $\tg$}

\begin{figure}

\epsscale{0.9}

%plotone{./figs/fig_gamma_amax_molcloud.eps}
\plotone{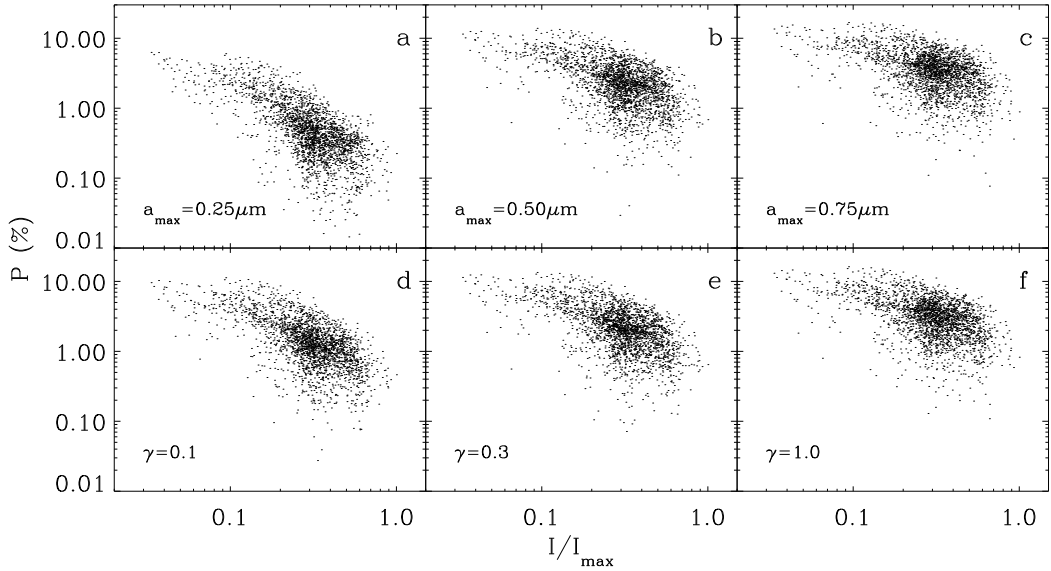}
\caption{P-I plots for the model molecular cloud obtained from the parameter study. The top row (a-c) shows how varying the upper grain-size cutoff, $a_{max}$, changes the P-I relation. Plot $b$ is in fact the result for our default molecular cloud. In the bottom row (d-f) we have replaced the real, spatially varying values for the radiative anisotropy, $\gamma$, with constant values $\gamma=0.1,\,0.3$ and $1.0$.\label{fig:poldeg_changing_dipole_mc}}

\epsscale{1.0}

\end{figure}

The top panel of Figure \ref{fig:poldeg_changing_dipole_mc} shows
the effects of changing $a_{max}$ over the range $[0.25,0.75]\mu$m.
Once again, a larger $a_{max}$ leads to a larger polarization degree
and flatter $P-I$ relation in bright sub-mm regions. Even the modest
change from $a_{max}=0.25\mu$m to $0.5\mu$m leads to a several-fold
increase in $P.$ The bottom panel in Figure \ref{fig:poldeg_changing_dipole_mc}
shows the effects of imposing a uniform anistropy, $\gamma=0.1,\,0.3$
or $1.0.$ As expected, the assumption of uniformity and the value
adopted for $\gamma$ has less of an effect than the choice of $a_{max}.$
Once again, assuming that the grains responsible for the $850\mu$m
polarized emission emit isothermally at 10K introduces only a small
error, as seen in Figure \ref{fig:poldeg_tempconst_histo}.

\section{SPECTRUM AND ALIGNMENT OF POLARIZATION\label{sec:Spectrum-and-alignment}}

In this section we consider two different types of spectra formed
from the polarized emission; first a spectrum of the polarized intensity,
and second, a spectrum of the polarization degree. The latter is known
as the `polarization spectrum', the form of which is highly indicative
of the source of the polarized emission. It helps discriminate between
populations of warm and cold aligned grains, which may, for example,
imply the presence of additional grain alignment via localised star
formation. 

\begin{figure}

%plotone{./figs/fig_pol_emission_spectrum_both.eps}
\plotone{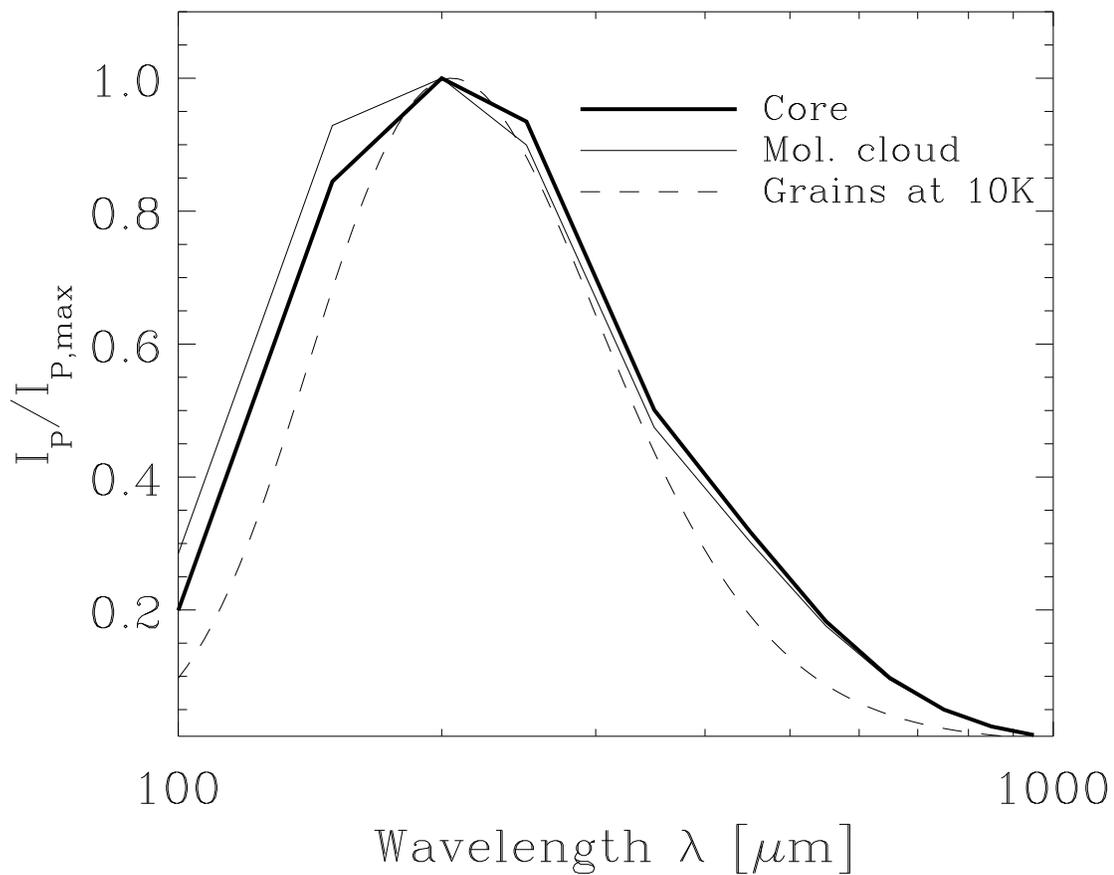}
\caption{The normalized spectrum of the map-averaged polarized intensity, $I_{P}$, from our core (\textit{thick solid line}) and molecular cloud (\textit{thin solid line}) models.  Also shown is a curve for grains with emission efficiencies $\propto \lambda^{-2}$ emitting at a temperature of 10K (\textit{dashed line}.)\label{fig:pol_emission_spectrum}}

\end{figure}

The map-averaged spectrum of the polarized intensity is presented
in Figure \ref{fig:pol_emission_spectrum}. The polarized intensity
deviates very little from the spectrum obtained with a population
of large grains emitting at $T=10$K, which is approximately consistent
with the mean temperature for the largest aligned grains (see Figures
\ref{fig:tg_dist} and \ref{fig:poldeg_tempconst_histo}). In our
models the grains exhibit a range of temperatures ($\Delta T_{g}\sim4$K)
which explains why our curve is broader than the isothermal 10K curve.
In our molecular cloud model, the aligned grains are smaller and therefore
also wamer on average than those responsible for the polarized emission
in the model core. As a result the polarized intensity spectrum for
the molecular cloud is shifted towards slightly shorter wavelengths,
although the effect is rather subtle. As was shown in section \ref{sub:Influence-of-gamma}
the small temperature fluctuations along a random line of sight have
very little effect on the emergent polarization degree. Instead the
temperature range merely acts to broaden the polarized intensity spectrum. 

\begin{figure}

%plotone{./figs/fig_pol_spectrum_both.eps}
\plotone{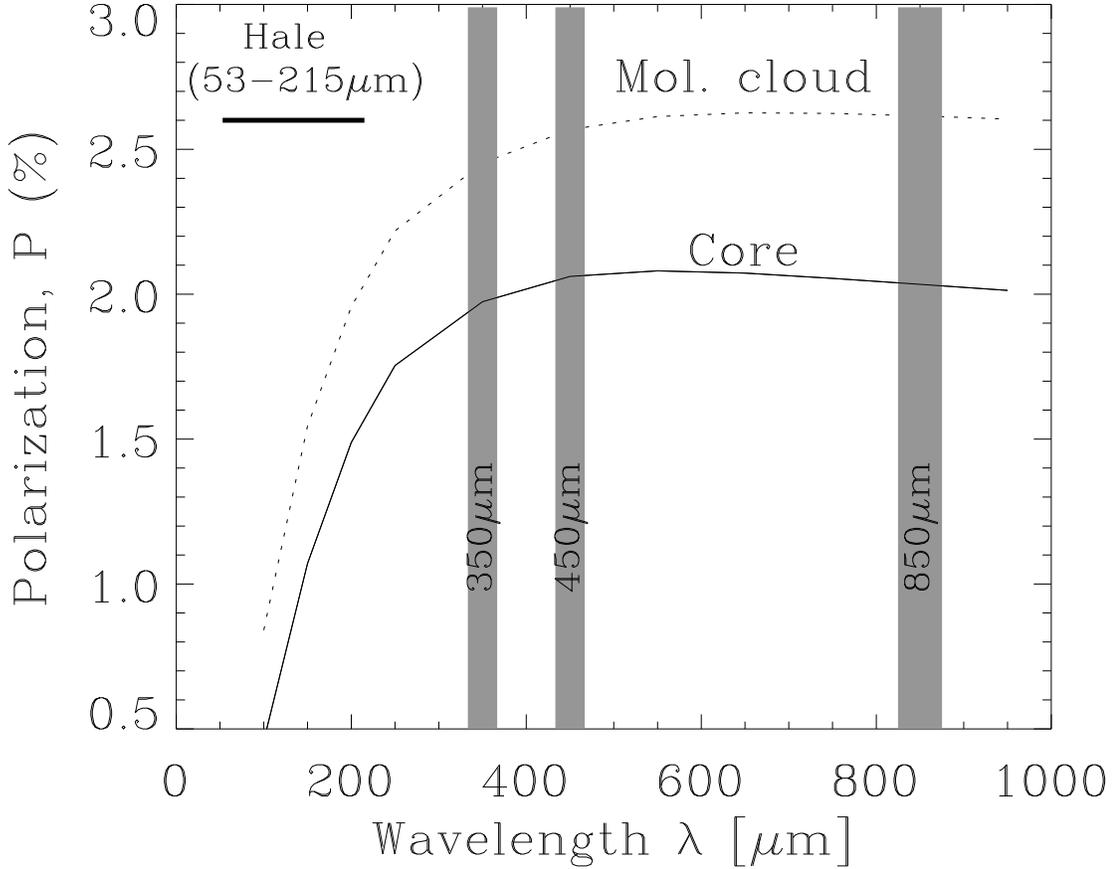}
\caption{The polarization spectra of the model core (\textit{solid line}) and molecular cloud (\textit{dashed line}), overlaid with the "atmospheric windows" of relatively weak atmospheric water absorption centered at $350,\,450$ and $850\mu$m (\textit{shaded regions}). The projectd Hale polarimeter waveband coverage is also shown. To calculate the polarization degree at a particular wavelength the total polarized emission, $I_{P}$, is integrated over the entire map and  divided by the total map emission.\label{fig:pol_spectrum}}

\end{figure}

The map-averaged degree of polarization as a function of wavelength
(the polarization spectrum) for both our models are shown in Figure
\ref{fig:pol_spectrum}. Superposed are the wavebands frequently observed
from the ground ($350,\,450$ and $850\mu$m), as well as the projected
wavelength range for the Hale polarimeter (promising a wavelength
coverage of $53-215\mu$m, Hildebrand, private communication), one
of the polarimeters proposed for SOFIA. For wavelengths longer than
$\sim400\mu$m the mean polarization degree is constant at $\sim2-2.5$\%,
dropping precipitously at wavelengths less than $350\mu$m where the
small, unaligned grains begin to dominate the sub-mm emission.

\subsection{Alignment of polarization and magnetic field\label{sub:Alignment-of-polarization}}

\begin{figure}

%plotone{./figs/fig_costheta_Nh_850_both.eps}
\plotone{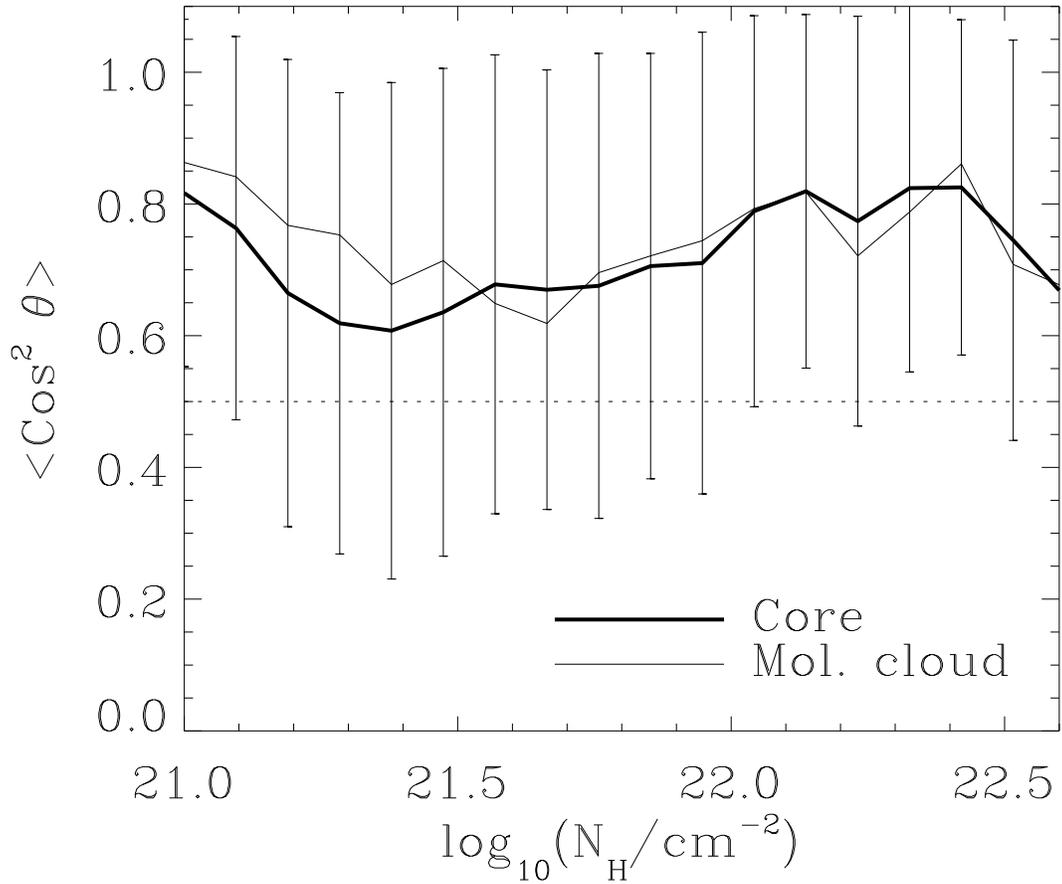}
\caption{The mean squared cosine of the angle between polarization and projected magnetic field vectors, $<\cos^2\theta>$, as a function of column density.  The error bars represent the r.m.s scatter about the mean. The polarization vectors at the default wavelength of 850$\mu$m are used although there is little wavelength dependence of these quantities.\label{fig:costheta_nh}}

\end{figure}

\begin{figure}

%plotone{./figs/fig_costheta_wav.eps}
\plotone{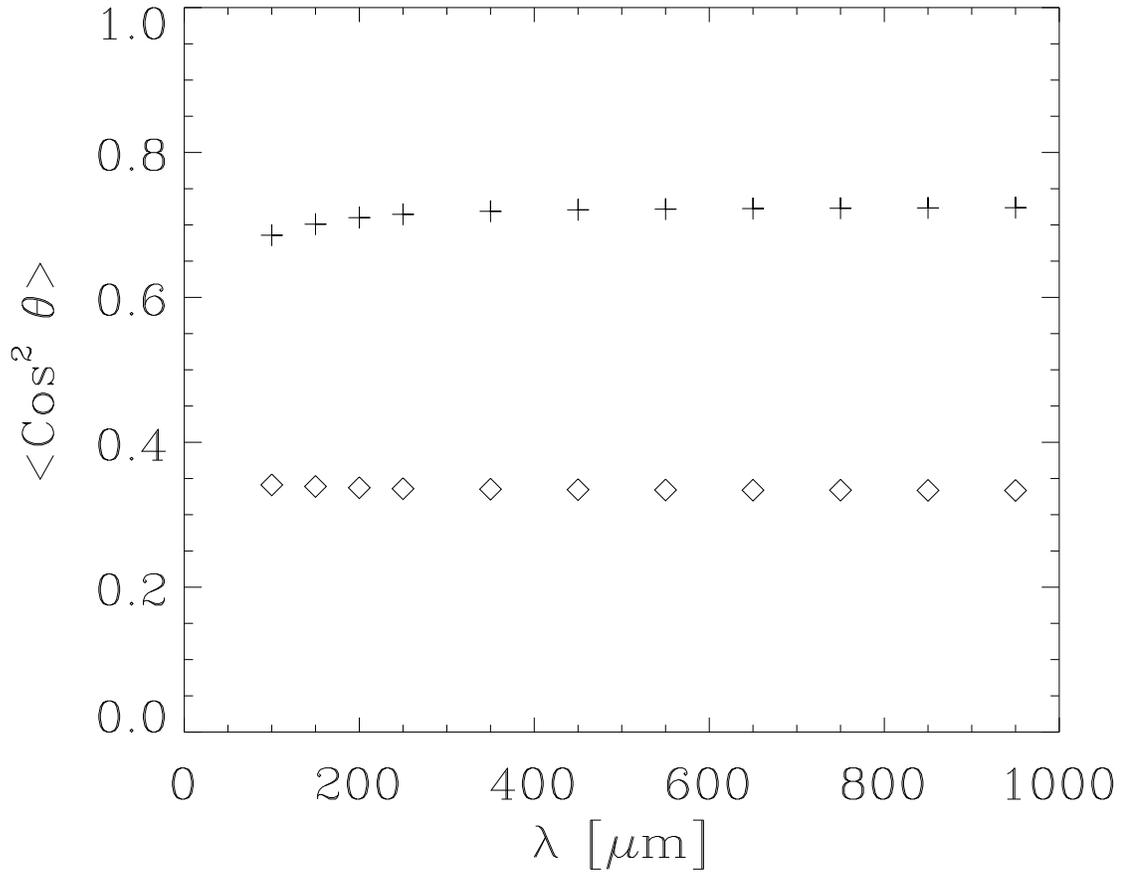}
\caption{The mean squared cosine of the angle between polarization and projected magnetic field vectors, $<\cos^2\theta>$, Averaged over maps of our model core made at various wavelengths (\textit{crosses}). The \textit{diamonds} represent the root mean squared (\textit{rms}) scatter about $<\cos^2\theta>$.  The results for the molecular cloud model are almost identical and are not shown.\label{fig:costheta_lambda}}

\end{figure}

The angle $\theta$ between the polarization and projected magnetic
field vectors is an indicator of the ability of polarization maps
to reveal the projected structure of the underlying magnetic field.
We obtain $\cos^2\theta$ from the dot product of the polarization and
projected magnetic field vectors,

\begin{equation} \cos^{2}\theta=\left(\mathbf{\frac{P\cdot B_{proj}}{|P||B_{proj}|}}\right)^{2},\label{eq:costheta}\end{equation}\\where
$\mathbf{B}$ is the mass weighted projected magnetic field. An informative
quantity is the degree of alignment as a function of column density.
The most natural way to express the degree of alignment is through
the mean $\cos^{2}\theta$, since this appears in the conventional
definition of the Rayleigh reduction factor (see Section \ref{sec:GRAIN-ALIGNMENT}).
In Figure \ref{fig:costheta_nh} the mean quantity $<\cos^{2}\theta>$
is shown as a function of $N_{H}$ for both our models. Although these
results are calculated at the default wavelength ($850\mu$m), the
following results do not change significantly with wavelength. The
alignment is approximately constant, $<\cos^{2}\theta>\sim0.75,$
throughout the range of column densities, although there is a large
scatter about this average value. If $\pvec$ and $\bvec$ were randomly
oriented with respect to one another we would expect $<\cos^{2}\theta>=0.5$.
From these results it seems that the polarization vectors do on average
trace the projected magnetic field vectors over the range of column
densities. The map-averaged $\cos^{2}\theta$ is shown as a function
of wavelength in Figure \ref{fig:costheta_lambda}. There is no significant
dependence on wavelength from $\lambda=100-1000\mu$m.

\section{Discussion\label{sec:Discussion}}

By considering very clumpy models for prestellar cores and molecular
clouds we found that:

1. Detectable levels of polarization ($P>1$\%) should be present
from clumpy, optically thick $(\av\sim10)$ prestellar cores and turbulent
molecular clouds. The alignment of large grains $(a\sim1\mu$m) via
the mechanism of radiative torques is shown to be an effective process
in our models, even though the only source of radiation is the mean interstellar radiation field. However, the presence of sufficiently large grains
is essential for this mechanism to work efficiently. 

2. The inclusion of clumpiness and a detailed solution of the radiative
transfer equation does not significantly change the qualitative nature
of the results described in CL05. In addition, we note that the distribution
in temperatures of the large, aligned grains is very narrow and the
assumption of isothermality amongst these grains introduces only a
negligibly small error into the polarization calculation. 

3. The emergent degree of polarization is only moderately sensitive
to simplifying assumptions about the radiative anisotropy. However,
it is extremely sensitive to the adopted upper cut-off of the power-law
distribution of grain sizes.

4. Despite the highly complex magnetic field geometry in our clumpy
models, the polarization vectors trace the mass-weighted projected
magnetic field vectors reasonably well, even for lines of sight where
$\av\sim10$.

5. The synthetic polarization spectra are flat longward of $\lambda\sim300\mu$m
but fall precipitously shortward of this wavelength. This is in part due to an absence of emission at these short wavelengths.  Multi-frequency
polarimetric studies as well as new models are necessary to further test the theory of grain alignment via radiative torques.

Many extensions to this work are also possible; for example, the eventual
onset of star formation in our core will introduce local sources of
highly anisotropic radiation.
These stars will illuminate the dense clumps from within, where grain alignment
from the ISRF alone might prove inadequate. The polarization due to
embedded stars will typically produce a different polarization spectrum,
introducing a warm component extending to smaller
grain sizes. The detailed effects of embedded star-formation are to be considered in a future paper.  Indeed, observed polarization spectra show an increase in polarization degree shortward of $300\mu$m, which can be most easily attributed to an additional populaiton of warm, aligned grains.  Since we only include the basal ISRF our calculations we lack both sufficiently warm grains and the diversity of conditions that prevail in real objects such as M17 and OMC-1 (Vaillancourt 2002).  For these reasons we do not expect our current models to exactly reproduce the observations shortward of 300$\mu$m.

Given that appreciable polarization signals arise in even our conservative models, it seems reasonable to assume that a measurable
polarization should be found throughout the lifetime of a core. The
polarization degree arising from the radiative torque mechanism is
determined by the largest grains, and is therefore sensitive to the
upper grain-size limit $a_{max}.$ If constraints can be placed on
the ambient radiation field then the polarization degree could be
an effective probe of the large grains in the interstellar dust population.
Such favorable conditions should prevail in the vicinity of individual
or compact clusters of stars where the radiative anisotropy is approximately
unity.  

Due to poor sub-mm wavelength coverage, it is currently difficult
to draw strong conclusions about the physical conditions in a cloud
from the observed polarization spectra (e.g. Hildebrand et al. 2000).
However, a number of sub-mm telescopes and detectors are planned for
the near future (e.g. Planck, Herschel, SOFIA), many of which will
eventually have polarimetric capabilities. In particular, SOFIA (the
Stratospheric Observatory for Infrared Astronomy), once equipped with
a polarimeter such as Hale, will dramatically enhance our ability
to map the magnetic field structure in prestellar cores, as well as
a variety of other environments. Measurements shortward of $300\mu$m
in a diverse sample of objects should prove particularly useful, since it is here that our models
predict a rapid decline in the polarization degree. Explaining the discrepancy between our models and observations in this spectral regime will be a strong test of the radiative torque theory.  Thus, the need for
spectropolarimetry is clear: not only is a broad wavelength coverage
required in order to describe the polarization spectrum, but spectral
information is essential for disentangling the various physical environments
that may exist along any given line of sight (Chen 1990, Xie et al.
1991, Hildebrand et al. 1999). Furthermore, our paper shows that a reliable statistic of the magnetic field topology can
be obtained via sub-mm polarimetry. Comparing these predictions with observations
will provide a valuable test for the grain alignment theory. As a consequence of 
such testing we will not only gain insight into the physics of star-formation but also gain confidence in future
studies of magnetic fields in circumstellar environments,
comet coma and tails, and other galaxies (see Lazarian 2003).

While in the above work we disregard both paramagnetic alignment
pioneered by Davis-Greenstein (1951) and mechanical alignments pioneered by Gold (1952), we believe that our estimates above will not be substantially
 altered if we include these mechanisms. As discussed in 
Lazarian et al. (1997) the conditions within the molecular clouds are close
to equilibrium ones. Thus without radiative torques the grains rotate 
essentially thermally. As the gas and grain temperatures are not very different
in molecular clouds, the paramagnetic alignment is marginal, even if grains
have superparamagnetic properties (see Roberge \& Lazarian 1999). The
relative gas grain velocities that are necessary for the
mechanical alignment are dominated by the gyroresonance acceleration (Yan
\& Lazarian 2003). For molecular clouds the resulting velocities are not
high enough, however (see Yan, Lazarian \& Draine 2004) to produce
efficient mechanical alignment (see Roberge, Hanany \& Messinger 1995,
Lazarian 1997 and references therein).     

After this paper was written we learned about the interesting and complementary work of Pelkonen
et al. (2006), which also extends the CL05 model by applying it to numerical simulations.
In contrast to our work, Pelkonen et al. study much smoother cores in the context of their parental cloud.  We restrict ourselves to one isolated core model, but calculate its polarized sub-mm emission in great detail. Pelkonen et al. assume a constant degree of anisotropy of $\gamma=0.7$ (e.g. Draine \& Weingartner 1996) which is a factor of two larger than our mass-averaged value.  In the context of their smooth cores this approximation may prove to be adequate, although in our calculations the properties of the anisotropy are rather different.  Despite our detailed calculations it is still not clear how to prescribe approximate values for $\gamma$ on the basis of the density structure alone. Despite these uncertainties, Pelkonen et al. also show, as we have done in this paper, that the observed $P-I$ relations can be reproduced within the radiative torque framework.

 Interestingly, Pelkonen et al. directly employ the empirical formula for the radiative torque given by equation \ref{eq:cl05}.  This is possible if the intracore radiation field can be described by a single effective extinction $A_{V}$.  Their approach is therefore a direct application of the CL05 method to 3D cloud structures. In our model, because of its extreme clumpiness, the radiation field cannot be described by a single parameter (like $A_{V}$) and requires detailed calculation. Unfortunately it seems unlikely that a simple empirical formula for the radiative torque exists for highly clumpy objects such as ours. Despite their differences these two complementary approaches represent the first attempts at developing the radiative torque mechanism for inhomogeneous media, and there is still much interesting work to be done in this area.

\acknowledgements{TJB is happy to acknowledge support provided by the NASA ATP program and NSF grant (AST 0507367). AL acknowledges the support by the NSF grants AST 0098597, AST 0243156 and AST 0507164, as well as by the
NSF Center for Magnetic Self-Organization in Laboratory and Astrophysical 
Plasmas, AC acknowledges the support by the NSF grant AST 0507164.}

%SOFIA (747)

%Herschel (polarimetry? space)

%SMA (Goodman interf.)

%ALMA (interf.)

%BIMA (interf.)

%Planck (space)

%%%%%%%%%%%%%%%%%%%%%%% Core results

%%%%%%%%%%%%%%%%%%%%%%%%%% results for Molecular cloud

%%%%%%%%%%%%%%%%%%%%%%%%%% Polarization spectra

\end{document}